\documentclass[twocolumn,showpacs,preprintnumbers,amsmath,amssymb]{revtex4}
\usepackage{graphicx}

\def\gtrsim{\mathrel{\mathpalette\vereq>}}

\begin{document}
\preprint{      
        \vbox{\hbox{SLAC-PUB-9650}},
        \vbox{\hbox{UCB-PTH-03/03}},
        \vbox{LBNL-52194}
        \vbox{hep-ph/0302131},}
\title{WMAPping out Neutrino Masses}
\author{Aaron Pierce}
\affiliation{
Theoretical Physics Group, Stanford Linear Accelerator Center, \\
Stanford University, Stanford, California, 94309, USA}
\author{Hitoshi Murayama}
\affiliation{Department of Physics,
        University of California,
        Berkeley, California, 94720, USA}
\affiliation{Theory Group,
Lawrence Berkeley National Laboratory,
Berkeley, California, 94720, USA}

\date{\today}
\begin{abstract}
  Recent data from from the Wilkinson Microwave Anisotropy Probe
  (WMAP) place important bounds on the neutrino sector.  The precise
  determination of the baryon number in the universe puts a strong 
  constraint on the number of relativistic species during Big-Bang 
  Nucleosynthesis.  WMAP data, when combined with the 2dF Galaxy 
  Redshift Survey (2dFGRS), also directly constrain the absolute mass 
  scale of neutrinos.  These results impinge upon a neutrino oscillation 
  interpretation of the result from the Liquid Scintillator Neutrino 
  Detector (LSND).  We also note 
  that the Heidelberg--Moscow evidence for neutrinoless double beta decay 
  is only consistent with the WMAP+2dFGRS data for the largest values of 
  the nuclear matrix element.
\end{abstract}
\pacs{14.60.Pq, 98.80.-k}
\maketitle
\setcounter{footnote}{0}
\setcounter{page}{1}
\setcounter{section}{0}
\setcounter{subsection}{0}
\setcounter{subsubsection}{0}


\section{Introduction}
Evidence for neutrino oscillation has steadily
mounted over the last few years, culminating in a picture that
presents a compelling argument for finite neutrino masses.  The
observation of a zenith-angle dependent deficit of $\nu_{\mu}$ from
cosmic ray showers at Super-Kamiokande \cite{SuperK}, provided strong
evidence for oscillations in atmospheric neutrinos.  Recent results on
solar neutrinos at the Sudbury Neutrino Observatory (SNO) \cite{SNO}
and reactor neutrinos at the KamLAND experiment \cite{KamLAND}, have
shed light on the solar neutrino problem.  These experiments have
provided strong evidence that the solar neutrino problem is solved by
oscillations corresponding to the Large Mixing Angle solution
\cite{MSW}.  Although clear oscillation data now exist in atmospheric,
reactor, and solar neutrino experiments, it remains to determine the
significance of the result from the Liquid Scintillator Neutrino
Detector (LSND) \cite{LSND1, LSND2}, which claimed evidence for
conversion of $\bar{\nu}_{\mu}$ to $\bar{\nu}_{e}$ with a $\Delta
m^{2}_{\nu}$ of order 1 eV$^2$.

While these extraordinary advances in experimental neutrino physics
were occurring, a concurrent revolution in experimental cosmology took
place.  Ushered in by the Boomerang, MAXIMA, and DASI  
measurements of the acoustic peaks in
the Cosmic Microwave Background (CMBR)~\cite{CMBR}, an era has begun 
wherein it is possible to make measurements of cosmological parameters with
previously unimaginable precision.  Most recently, the striking data
\cite{WMAP1} from the Wilkinson Microwave Anisotropy Probe (WMAP) have
vastly improved our knowledge of several fundamental cosmological
parameters \cite{WMAP2}.  Because cosmology would be significantly
affected by the presence of light species with masses of order 1 eV,
the new WMAP data strongly constrain neutrino masses in this range.
We will show this brings cosmology into some conflict with the LSND
result in two ways.

First, WMAP determines the baryon to photon ratio very 
precisely. This removes an important source of uncertainty in the 
prediction of Big-Bang Nucleosynthesis (BBN) for the primordial 
abundance of ${ }^{4}$He.  This allows for a strong limit to be 
placed on the number of relativistic species present at BBN, disfavoring the
LSND result.
Secondly, WMAP, when combined with data from the 2 degree Field 
Galactic Redshift Survey (2dFGRS) \cite{2dFGRS}, CBI \cite{CBI}, 
and ACBAR \cite{ACBAR}, is able to place stringent 
limits on the amount that neutrinos contribute to the critical density 
of the universe.  This second constraint results in an 
upper mass-limit on neutrinos that 
contradicts the LSND result in all but one ``island''
of parameter space not ruled out by other experiments.  The second 
constraint also impinges on the recent evidence 
for neutrinoless 
double beta decay from the Heidelberg-Moscow experiment
\cite{Klapdor-Kleingrothaus}. 
\section{The LSND Result}
The LSND experiment used decays of stopped 
anti-muons at the LAMPF facility (Los Alamos) to look for the appearance of
anti-electron-neutrinos.  They reported the oscillation
probability $P(\bar{\nu}_\mu \rightarrow \bar{\nu}_e) = (0.264 \pm
0.067 \pm 0.045)\%$, representing a 3.3$\sigma$ signal.

If the result at the LSND experiment were a true indication of
oscillations, it would have profound implications for our
understanding of neutrinos.
Solar and atmospheric neutrinos have already determined two
neutrino mass-squared differences to be $\Delta m^{2}_{solar} \sim
10^{-4}$ eV$^{2}$ and $\Delta m^{2}_{atm} \sim 10^{-3}$ eV$^{2}$.
However, taking into account the Bugey exclusion region \cite{Bugey},
the LSND experiment points to a mass difference (see Figure 1) $\Delta
m^{2}_{LSND} > 10^{-1}$ eV$^{2}$.  The presence of this completely
disparate mass difference necessitates the introduction of a fourth
neutrino~\footnote{This statement assumes CPT. If neutrinos and
  anti-neutrinos have different mass spectra, it may still be possible
  to accommodate LSND together with solar, reactor, and atmospheric
  neutrino data within three generations alone~\cite{CPTviolation}.
  However, see also \cite{Strumia}.}.
Because LEP has determined the number of active neutrino species to be
three, this fourth neutrino must be sterile, having extraordinarily
feeble couplings to the other particles of the standard model.

\begin{figure}[t]
\includegraphics[width=\columnwidth]{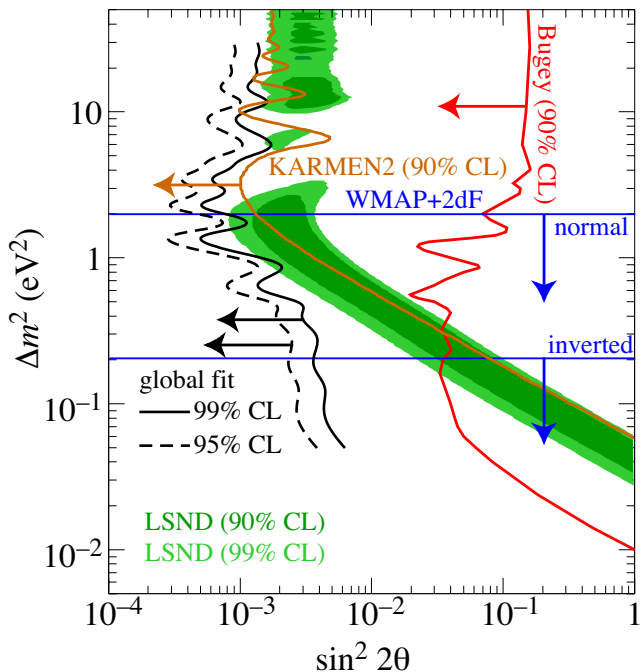}\label{fig:global}
\caption{The LSND Allowed region, with Bugey and Karmen~\cite{Karmen} 
  exclusion regions.
  The constraints from the global fit \cite{global} as well as the
  limit of~\cite{Hannestad} from 
  the combination of WMAP and 2dFGRS data are also shown.  There are 
  two lines, corresponding to the $3+1$ (normal) and $1+3$ (inverted)  
  spectra.
  The contours from the global fit would, of course, continue on to 
  lower values of $\Delta m^{2}$, but Ref.~\cite{global} did not show 
  this region.}
\end{figure}

The introduction of this fourth neutrino species results in principle
in two characteristic types of spectra, $2+2$ and $3+1$.  Two sample
spectra of these types are shown in Figure 2.

\begin{figure}[t]
\includegraphics[width=\columnwidth]{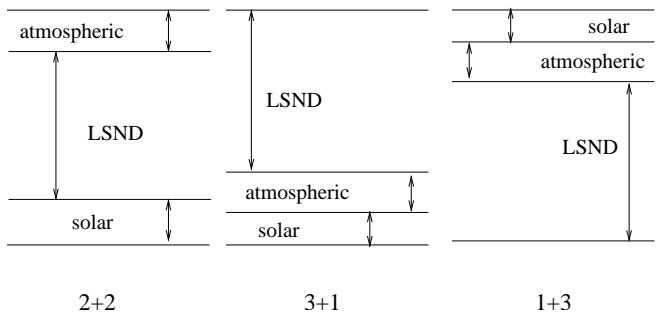}\label{fig:spectra}
\caption{Sample neutrino spectra in the light of LSND.  Different
permutations are also possible.}
\end{figure}
However, recent results from SNO \cite{SNO} and Super-Kamiokande
\cite{SuperKactive}, have indicated that the oscillations responsible
for the atmospheric and solar neutrino anomalies involve transitions
primarily between active neutrinos.  This means that it is difficult
to put the sterile part of the neutrino in either the solar or
atmospheric pair in the $2+2$ spectrum.  A recent quantitative
analysis \cite{Strumia} found this $2+2$ spectrum to be completely
ruled out, while a $3+1$ spectrum was allowed at the $99\%$ confidence
level \cite{global, Strumia}.  The tension for the 
$3+1$ spectrum is in large part due to the
lack of a signal in short-baseline disappearance experiments such as
CDHSW \cite{CDHS} and Bugey.  Adding additional sterile neutrinos can
only marginally improve this agreement \cite{Smirnov}.  
In the next two sections, we show how this allowed window is further 
constrained by cosmological considerations. 
\section{Big-Bang Nucleosynthesis}
By measuring the primordial abundance
of ${ }^{4}$He, one can place bounds on extra relativistic degrees of
freedom at the time of Big Bang Nucleosynthesis (BBN) \cite{Hoyle}.
These bounds are usually quoted in terms of a number of effective
allowed neutrino species, $N_{\nu}^{\it eff}$.  Additional degrees of
freedom tend to increase the expansion rate of the universe, which
causes neutrons to freeze out at an earlier time, at a higher
abundance.  This abundance translates into more primordial 
${ }^{4}$He for a given baryon to photon ratio, $\eta$.  Therefore,
knowledge of primordial ${ }^{4}$He abundance along with a separate
determination of $\eta$ places a bound on $N_{\nu}^{\it eff}$.  On the
other hand, for a fixed $N_{\nu}^{\it eff}$, a higher $\eta$ results in a
higher abundance for primordial ${ }^{4}$He; so, incomplete knowledge
of $\eta$ degrades the constraint on $N_{\nu}^{\it eff}$.

In the era before precise CMBR measurements, BBN data alone were
utilized to set the bound.  Measurements of primordial deuterium or
lithium were used for the separate determination of $\eta$.  An
aggressive analysis by \cite{OliveBBN} cited a limit of $N_{\nu}^{\it eff}
< 3.4$ at 2$\sigma$, and consequently found that LSND data were
strongly disfavored by BBN \cite{Olive}.  However, the data for
primordial light element abundances were somewhat muddled, with some
measurements of lithium and deuterium preferring substantially lower
values of $\eta$ than others. Due to the presence of these data, a
conservative bound $N_{\nu} < 4$ was often taken \cite{Lisi}.  In
fact, using lithium data alone, \cite{OliveThomas}, found that even
$N_{\nu}^{\it eff} =4.9$ was acceptable at the 95\% confidence level.

However, after precise measurements of the CMBR, the situation has
changed.  The WMAP experiment has determined \cite{WMAP2} $\Omega_{b}
h^{2} = 0.224 \pm 0.001$, corresponding to an $\eta =
6.5^{+0.4}_{-0.3} \times 10^{-10}$.  For the central value above, the expected 
${ }^{4}$He abundance, $Y_{p}$, is roughly
$Y_{p}=0.249+0.013(N_{\nu}^{\it eff}-3)$.  The status of primordial Helium
measurements remains controversial.  One helium measurement quotes a
value $Y_{p} =0.244 \pm 0.002$ \cite{HeHigh}, while another quotes
$Y_{p} =0.235 \pm 0.002$ \cite{HeLow}.  To deal with the discrepancy in
these measurements, the Particle Data Group (PDG) assigns an
additional systematic error, taking $Y_{p}=0.238 \pm 0.002 \pm
0.005$ \cite{PDG}.  To be completely conservative, we will take the
higher helium abundance, and assign to it the additional systematic
error of the PDG, namely, we take $Y_{p}=0.244 \pm 0.002 \pm 0.005$.
Using the formulae of \cite{turnerlopez} for ${ }^{4}$He in terms of
$N_{\nu}^{\it eff}$ and $\eta$, we find $N_{\nu}^{\it eff} < 3.4$ at the 95\%
(two-sided) confidence level, leaving no room for the extra neutrino of LSND.
Using the only slightly less conservative approach of adopting the PDG
central value and error, we find $N_{\nu}^{\it eff} < 3.0$ at the 95\%
(two-sided) confidence level.
  
Of course, additional systematic errors in the helium abundance
measurements may be found.  The fact that 3 neutrinos is barely
consistent at the 95\% confidence level might cause some suspicion
that there are unknown systematics at work.  However, to get
$N_{\nu}^{\it eff}=4$ at the 95\% level would require inflating the errors
on the PDG central value dramatically, to $Y_{p}=0.238 \pm 0.011$.

It is possible that an asymmetry in the leptons could effectively
prevent the oscillation into sterile neutrinos \cite{Foot}.  
We find
that a large pre-existing asymmetry of $L^{(e)} \sim 10^{-2}$ would be
sufficient to suppress the production of sterile neutrinos below the
BBN constraint.  Here, $L^{e}$, represents the total asymmetry felt by
electron neutrinos, 
$L^{e} = 2 L_{\nu_{e}} + L_{\nu_{\mu}} + L_{\nu_{\tau}}$, 
with $L_{\nu_{i}}=(N_{\nu_{i}}-N_{\nu_{j}})/N_{\gamma}$.  
Smaller $L^{e}$ (as low as $\sim 10^{-5}$) can suppress sterile neutrino
production, but oscillations tend to erase asymmetries of this size. 
While a lepton asymmetry of $10^{-2}$
size does not bias the rate for the processes such 
as $n +e^{+} \rightarrow p + \bar{\nu}_{e}$ significantly enough to affect 
BBN, an asymmetry as large as $10^{-1}$ would.  CMBR
constraints also cannot exclude the possibility of a lepton asymmetry of
$10^{-2}$, so this possibility can not be excluded.  
It does not appear that neutrino
oscillations themselves can create this asymmetry \cite{DiBari}.  One 
has to assume that the asymmetry existed before the BBN,
possibly generated by a mechanism similar to that 
in \cite{March-Russell:1999ig}.

\section{Weighing Neutrinos with Large Scale Structure}
WMAP has provided an additional constraint
on LSND.  As noted, for example, in \cite{Hu}, Galactic Surveys
provide a powerful tool to constrain the masses of neutrinos.
Neutrinos decouple at temperatures well above those at which structure
forms.  They then free-stream until they become non-relativistic.
This tends to smooth out structure on the smallest scales.  On scales
within the horizon when the neutrinos were still relativistic, the
power spectrum of density fluctuations is suppressed as \cite{Hu}:
\begin{equation}
\frac{\Delta P_{m}}{P_{m}} \approx -8 \frac{\Omega_{\nu}}{\Omega_{m}}
\end{equation}
The 2dFGRS experiment used this fact to place a limit on the sum of neutrino 
masses: $\Sigma m_{\nu} < 1.8$ eV \cite{2dFGRS}.

Recent data from WMAP greatly improve this measurement.  A key
contribution is the fact that WMAP and 2dFGRS overlap in the
wavenumbers probed. This allows a normalization of the 2dFGRS power
spectrum from the WMAP data.  The WMAP satellite also precisely
determines $\Omega_{m}$.  Since depletion of power at small scales is
sensitive to the ratio of $\Omega_{\nu}/\Omega_{m}$, a more accurate
determination of $\Omega_{m}$ leads to a better bound on the neutrino
mass.  The ultimate result from combining data from 2dFGRS, ACBAR,
CBI, and WMAP is $\Omega_{\nu} h^{2} < 0.0076$ (95 \% confidence
level) \cite{WMAP2}.  
The bound on $\Omega_{\nu} h^{2}$ places the bound masses 
$m_{\nu} < 0.23$~eV (3 degenerate Neutrinos, 95\% 
confidence level). Note that using the WMAP data, \cite{Hannestad} finds a 
more conservative bound of 
\begin{equation} \label{eqn:3bound}
m_{\nu} < 0.33 \mbox{ eV (3 Degenerate Neutrinos, 95\% CL)}.  
\end{equation}
The primary difference in the bounds is that \cite{Hannestad} allows the 
bias factor to float in the analysis.  

In the case where there are four neutrinos, the bound on neutrino masses is 
somewhat relaxed.  As noted by \cite{OldHannestad}, the bound on $\sum m_{\nu}$
is anti-correlated with the value of the Hubble constant. On the other hand, 
limits on $N_{\nu}$ are correlated with Hubble constant.  Playing these two
effects against one another allows the weakening of bounds on $m_{\nu}$ for 
$N_{\nu}=4$.
For the $3+1$ spectrum shown in Figure 2, 
again allowing the bias parameter to float,  
\cite{Hannestad} finds a bound of 
\begin{equation} \label{eqn:4bound}
m_{\nu} < 1.4 \mbox{ eV (3+1 Neutrinos, 95\% CL)}.  
\end{equation}
This bound was derived assuming three degenerate
active neutrino species.  In the case where the neutrinos are not all 
degenerate, in principle one might expect the bound to 
by slightly modified, as the scale where free-streaming stops would be 
shifted.  In practice, however, this has only a very small quantitative 
effect~\cite{Steen}, so we negelect it in our discussion.  
Also, the above mass limit was placed assuming that the heavy neutrino has
standard model couplings.  These couplings determine when the neutrino
decouples from thermal equilibrium.  If the neutrino decoupled
sufficiently early, it might have been substantially diluted relative
to the active neutrinos.  Consequently, it could contribute a
relatively small amount to the critical density today.  However, we do
not expect this to be the case for an LSND neutrino.  While one must
be careful to take into account plasma effects, \cite{Notzold}, that might
keep sterile neutrinos out of equilibrium at high temperatures, these
become negligible in time for LSND neutrinos to thermalize before decoupling. 
Reference~\cite{DiBari}
found that an additional sterile neutrino in a $3+1$ scheme
was nearly completely thermalized over the entire favored LSND mixing region.
Since the neutrino ultimately decouples at temperatures of order 10 MeV, 
abundance of these neutrinos will not be diluted by the entropy
produced at the QCD phase transition.  This assures us that the limit of
Eq.~(\ref{eqn:4bound}) is applicable for the heavy LSND neutrino as
well.  

Fitting the LSND result within a two neutrino oscillation picture
requires (see Figure 1) a neutrino mass greater than the square-root
of smallest allowed $\Delta m^{2}$.  This gives $m_{\nu} \gtrsim 0.45$
eV.  Comparing this with the bound on the neutrino mass in the 3+1 
scheme, Eq.~(\ref{eqn:4bound}), one sees that the
minimum LSND result is significantly squeezed by the large scale
structure measurement alone.  Taking into account a full $3+1$
neutrino oscillation analysis, fully incorporating data from CDHSW and
Bugey, we are forced into the small angle portion of the LSND allowed
region.  This means higher masses.  At the 99\% confidence level,
the allowed region contains four islands, corresponding neutrinos with 
masses \cite{global} (See Fig. 1) 
\begin{equation}
m_{\nu} \gtrsim 0.9 \mbox{ eV, } 1.4 \mbox{ eV, } 2.2  \mbox{ eV, } 3.5 
\mbox{ eV}.   \end{equation} 
All but the first of these conflict with Eq.~(\ref{eqn:4bound}), though 
the second is marginal.  If, unlike the analysis of \cite{Hannestad}, 
one were to take a prior for the bias factor, the conflict would become 
stronger. So the LSND experiment is strongly constrained by large scale 
structure measurements alone.

If instead of the $3+1$ spectrum, we had chosen the inverted
$1+3$ spectrum, the conflict would have been sharper.  In the inverted case,
the bound coming from large scale structure is stronger, (see Fig. 1), and the
LSND islands are easily excluded.

It is interesting to note that the WMAP experiment also detected a
relatively early re-ionization period, $z_{reionize} \sim 20$.  This
implies an early generation of stars responsible for the energy of
re-ionization during this period.  Early star formation disfavors warm
dark matter, consistent with the above statements that neutrinos make
up a small fraction of the critical density.
\section{Neutrinoless Double Beta Decay}
The limit on the neutrino mass
from the combination of WMAP and 2dFGRS data is also interesting in the
context of the neutrinoless double beta decay.  The Heidelberg--Moscow
experiment claimed a signal of neutrinoless double beta decay
\cite{Klapdor-Kleingrothaus}, which would indicate that neutrinos have
Majorana masses.  The relevant neutrino mass for the signal is the
so-called effective neutrino mass $\langle m_\nu \rangle_{ee} = |\sum_i
m_{\nu_i} U_{ei}^2|$.  The nuclear matrix elements 
in~\cite{Klapdor-Kleingrothaus} lead to the preferred range 
$\langle m_\nu \rangle_{ee} = (0.11$--$0.56)$~eV, while the reanalysis 
in \cite{Vogel} gives 0.4--1.3~eV using a different set of nuclear matrix 
elements.  This result does not require the presence of an additional 
(sterile) neutrino species, so the BBN limits need not apply.  However, 
this high value of $\langle m_\nu \rangle_{ee}$ together 
with solar, reactor, and atmospheric neutrino data on mass splittings, 
require the three neutrinos to be nearly degenerate.  In this case, the 
three degenerate neutrino bound of 
Eqn.~\ref{eqn:3bound} is appropriate, and the WMAP+2dFGRS 
data would therefore require $m_{\nu_i} < 0.33$~eV, 
or $m_{\nu_i} < 0.23$~eV, if the prior is taken on the bias factor. This 
large scale structure limit excludes the deduced range of the effective 
neutrino mass in~\cite{Vogel} completely.  However, using the largest 
values of the nuclear matrix element in~\cite{Klapdor-Kleingrothaus}, 
a window is still allowed.  Also, a recent review of the evidence for 
neutrinoless double beta decay assigns a somewhat larger error for the 
matrix element, and the largest allowed values of the matrix element 
could correspond to an effective neutrino mass as small 
as 0.05 eV \cite{KlapdorReview}.  So, the WMAP+2dFGRS constrains the 
claimed evidence for the neutrinoless double beta decay, but this statement
is dependent on what is assumed about the nuclear matrix elements.  
Moreover, the WMAP+2dFGRS result has nothing to say about the 
Heidelberg-Moscow result if the neutrinoless double beta decay arises from
a source other than Majorana neutrinos, such as supersymmetric models
with $R$-parity violation \cite{RPV}.

\section{Conclusions} \label{sec:conclusion}
Recent precise cosmological measurements have
given strong indications against the presence of an additional sterile
neutrino in the range that would explain the LSND result.  Bounds from
BBN disfavor the presence of any additional neutrinos that do not decouple 
before the QCD phase transition.  Large Scale Structure
disfavors the presence of neutrinos with mass in the eV range.  

It seems difficult to reconcile
LSND with the cosmological data.   
We have already discussed the possibility of having a large pre-existing 
lepton asymmetry of $L \sim 10^{-2}$.  Another possibility is to have 
CPT violation.  In this case,
the BBN constraint disappears, because no new light species are introduced. 
In addition, the large scale structure constraint is ameliorated, as only an
anti-neutrino would need to be heavy, but not its CPT neutrino partner.  
However, KamLAND data, when taken in concert with data from Super--Kamiokande
may disfavor this possibility~\cite{Strumia}.
The neutrino mixing result of LSND will be tested directly at the
MiniBoone Experiment at Fermilab \cite{Miniboone}.  

We also note that the cosmological
data do not prefer the neutrinoless double beta decay in the mass
range claimed by Heidelberg--Moscow experiment, unless the nuclear matrix
element is very large.  

\acknowledgments{
A.~Pierce thanks L.~Dixon and M.~Peskin for
useful and stimulating discussions that led to the idea for this work,
as well as for their insightful comments on the draft.  Both of us
thank Kev Abazajian and Steen Hansen for useful discussions.  As revisions
to this paper were being completed, the revision version 
of \cite{Pasquale} appeared,
which came to similar conclusions regarding the viability of a scenario
with a large lepton asymmetry.  The work of
A.~Pierce was supported by the U.S. Department of Energy under
Contract DE-AC03-76SF00515. The work of H.~Murayama was supported in
part by the DOE Contract DE-AC03-76SF00098 and in part by the NSF
grant PHY-0098840.
}

\end{document}